\documentclass[sigconf]{acmart}
\AtBeginDocument{%
  }

\setcopyright{acmlicensed}
\copyrightyear{2025}
\acmYear{2025}
\setcopyright{acmlicensed}\acmConference[SaTS '25]{Proceedings of the 2025 Workshop on Security and Privacy of AI-Empowered Mobile Super Apps}{October 13--17, 2025}{Taipei, Taiwan (Province of China)}
\acmBooktitle{Proceedings of the 2025 Workshop on Security and Privacy of AI-Empowered Mobile Super Apps (SaTS '25), October 13--17, 2025, Taipei, Taiwan (Province of China)}
\acmDOI{10.1145/3733824.3764877}
\acmISBN{979-8-4007-1912-7/2025/10}

\usepackage[inline]{enumitem}
\usepackage{booktabs}
\usepackage[subtle]{savetrees}
\usepackage{stfloats}

\begin{document}

\title{Who Moved My Transaction? Uncovering Post-Transaction Auditability Vulnerabilities in Modern Super Apps}


\author{Junlin Liu}
\affiliation{%
  \institution{Peking University}
  \city{Beijing}
  \country{China}}
\email{jlinliu@pku.edu.cn}

\author{Zhaomeng Deng}
\affiliation{%
  \institution{Peking University}
  \city{Beijing}
  \country{China}}
\email{infinityedge@pku.edu.cn}

\author{Ziming Wang}
\affiliation{%
  \institution{Peking University}
  \city{Beijing}
  \country{China}}
\email{wangzim@stu.pku.edu.cn}

\author{Mengyu Yao}
\affiliation{%
  \institution{Peking University}
  \city{Beijing}
  \country{China}}
\email{mengyuyao@stu.pku.edu.cn}

\author{Yifeng Cai}
\affiliation{%
  \institution{Peking University}
  \city{Beijing}
  \country{China}}
\email{caiyifeng@pku.edu.cn}

\author{Yutao Hu}
\affiliation{%
  \institution{Huazhong University of Science and Technology}
  \city{Wuhan}
  \country{China}}
\email{yutaohu@hust.edu.cn}

\author{Ziqi Zhang}
\authornote{Corresponding author(s).}
\affiliation{%
  \institution{University of Illinois Urbana-Champaign}
  \city{Champaign}
  \state{IL}
  \country{USA}}
\email{ziqi24@illinois.edu}

\author{Yao Guo}
\affiliation{%
  \institution{Peking University}
  \city{Beijing}
  \country{China}}
\email{yaoguo@pku.edu.cn}

\author{Ding Li}
\authornotemark[1]
\affiliation{%
  \institution{Peking University}
  \city{Beijing}
  \country{China}}
\email{ding_li@pku.edu.cn}


\begin{abstract}
Super apps are the cornerstones of modern digital life, embedding financial transactions into nearly every aspect of daily routine. The prevailing security paradigm for these platforms is overwhelmingly focused on pre-transaction authentication, preventing unauthorized payments before they occur. We argue that a critical vulnerability vector has been largely overlooked: the fragility of post-transaction audit trails. We investigate the ease with which a user can permanently erase their transaction history from an app's interface, thereby concealing unauthorized or sensitive activities from the account owner. To quantify this threat, we conducted an empirical study with 6 volunteers who performed a cross-evaluation on six super apps. Our findings are alarming: all six applications studied allow users to delete transaction records, yet a staggering five out of six (83+\%) fail to protect these records with strong authentication. Only one app in our study required biometric verification for deletion. This study provides the first concrete evidence of this near-ubiquitous vulnerability, demonstrating a critical gap in the current mobile security landscape and underscoring the urgent need for a paradigm shift towards ensuring post-transaction audit integrity.
\end{abstract}



\maketitle

\section{Introduction}
In many digital economies, super apps are not merely applications but are foundational operating systems for daily life. They seamlessly integrate communication, social media, e-commerce, and, most critically, financial services, processing trillions of dollars in transactions annually~\cite{yang2023sok,yang2025understanding}. Consequently, securing these platforms is a matter of paramount importance.

The current security focus, however, exhibits a crucial blind spot. The industry has invested immense resources in a strategy that can be termed pre-transaction defense~\cite{zhang2023don,cai2025can}. This approach involves erecting sophisticated fortifications, such as biometric authentication and AI-driven fraud detection, with the explicit goal of stopping illicit payments before they happen. While essential, this raises a vital question: what happens if an unauthorized transaction gets through? Can we guarantee its discovery?

This paper argues that the integrity of the post-transaction audit trail is a dangerously neglected aspect of security. This record is a cornerstone of user trust and control, yet it can often be easily erased, allowing malicious actors or even family members to conceal their activities. To address this gap, we introduce the principle of Post-Transaction Auditability and provide the first empirical study to measure this vulnerability in dominant super apps. Our findings are alarming: while all six tested applications allow record deletion, a staggering five out of six do so without strong authentication.

The main contributions of this paper are:
\begin{itemize}[leftmargin=*,nosep]
    \item We are the first to identify and formalize the threat of post-transaction record tampering in super apps, presenting a realistic threat model that considers both opportunistic insiders and malicious software.
    \item We introduce Post-Transaction Auditability as a new, critical security principle for financial applications, shifting focus from merely preventing bad transactions to ensuring they can always be discovered.
    \item We present the first empirical study that quantifies this vulnerability across six dominant super apps, demonstrating that the lack of protection is a systemic and widespread issue (83+\% of tested apps are vulnerable).
\end{itemize}

\vspace{-1em}
\section{Background}

\noindent\textbf{Existing Research on Payment Security.} 
Extensive research in mobile payment security has concentrated on pre-transaction vulnerabilities. This body of work covers critical topics such as defeating biometric sensors~\cite{cai2024famos,wang2025born}, phishing attacks for credentials~\cite{kishnani2022privacy,tang2020all}, and analyzing risks in NFC and QR code protocols~\cite{giese2019security,klee2020nfcgate}. Similarly, UI security research has often focused on preventing tapjacking or other overlay attacks~\cite{zhou2020ui,malisa2017detecting}. However, the specific threat of a legitimate (or seemingly legitimate) user intentionally corrupting the audit trail after a successful transaction remains a significant and unaddressed research gap. Our work distinguishes itself by focusing on \textit{preserving the record}.

\noindent\textbf{Threat Model.} 
We formalize our investigation with a clear threat model: the attacker’s primary goal is to permanently hide one or more transaction records from the user interface, thereby preventing the legitimate account owner from detecting the activity. We consider two representative attacker profiles, i.e., an opportunistic insider or a malicious software.

\vspace{-1em}
\section{Empirical Study Demo}
\noindent\textbf{Methodology.} 
To assess the real-world prevalence of this vulnerability, we conducted a controlled empirical study designed to answer a specific research question: What level of security do leading super apps implement to protect their transaction records from user-initiated deletion? We selected six of the world's most dominant super apps. These platforms are representative of the ecosystem, each boasting hundreds of millions of daily active users and featuring a deeply integrated payment system. 

The study involved six volunteers, and was conducted in a controlled lab setting. We prepared six smartphones, each pre-configured with a test account for one of the target applications. Each volunteer was randomly assigned a device at the beginning of the experiment. We employed a cross-evaluation design where participants rotated among the devices, ensuring that every volunteer systematically evaluated all six applications. For each app, the designated task was to locate the most recent transaction record, attempt to delete it, and meticulously document the UI path and any authentication challenges encountered during the process.

\noindent\textbf{Results.} 
Our study revealed a critical and widespread security vulnerability regarding how leading super apps handle the deletion of financial records. The most alarming finding is the near-complete absence of strong authentication for this sensitive action, allowing for the easy concealment of transactions.

\begin{table}[h]
\centering
\caption{Authentication Requirements for Deleting Transaction Records in Six Dominant super apps.}
\label{tab:auth_results}
\vspace{-1em}
\begin{tabular}{ll|ll} 
\toprule
\textbf{App} & \textbf{Verification Level} & \textbf{App} & \textbf{Verification Level} \\ 
\midrule
AliPay       & Biometrics                  & WeChat        & Pop-up Only                 \\
Taobao        & Pop-up Only                 & JD        & Pop-up Only                 \\
TikTok        & Pop-up Only                 & MeiTuan        & Pop-up Only                 \\ 
\bottomrule
\end{tabular}
\vspace{-3em}
\end{table}

The detailed findings of our experiment are summarized in \autoref{tab:auth_results}. The central conclusion is that a staggering five out of six applications tested (approximately 83\%) permit users to permanently erase transaction histories without requiring a password, PIN, or biometric verification.

As shown in \autoref{tab:auth_results}, only a single application, AliPay, has implemented a robust biometric safeguard. The vast lack of protection demonstrates that the risk is not an isolated oversight but appears to be a systemic issue and the de facto standard for the majority of the market leaders we investigated. While the existence of a deletion feature in all six apps highlights user demand for such functionality, it has clearly been implemented at the expense of fundamental audit integrity.

\noindent\textbf{Findings.} 
In conclusion, the absence of strong verification mechanism for deleting financial records indicates a systemic failure in design philosophy. It suggests that a majority of super app developers prioritize UI convenience over the fundamental security principle of audit integrity. This is a dangerous trade-off that leaves billions of users exposed.

For users, this study highlights a hidden risk. The assumption that their transaction history is a secure and reliable record is false. For the industry, our work is an urgent call to action. Implementing strong authentication for record deletion is not a complex engineering challenge; it is a matter of acknowledging the risk and prioritizing user security.

Meanwhile, as a preliminary study, our sample size of six super apps and six volunteers is small. However, the result provides a strong signal that this is a widespread issue worthy of immediate attention and larger-scale investigation.

\section{Conclusions and Future Work}
This paper provides the first empirical evidence of a critical vulnerability in modern super apps: the lack of strong authentication for deleting post-transaction audit trails. Our finding that 83.3\% of dominant platforms are vulnerable decisively challenges the current pre-transaction security paradigm. The integrity of a financial record is a fundamental security requirement, not a convenience.

Building on this preliminary study, which provides a strong signal of a widespread issue, our future work will proceed in two directions. First, we will develop an automated testing pipeline to perform a comprehensive, large-scale market analysis. Second, we plan to formalize our criteria into a public ``Post-Transaction Auditability'' benchmark to empower developers and allow for standardized comparison of apps.

\begin{acks}
We would like to thank the anonymous reviewers for
their valuable feedback. 
This work was supported by the National Science and Technology Major Project of China (2022ZD0119103). 
\end{acks}

\bibliographystyle{ACM-Reference-Format}
\bibliography{ref}


\end{document}